\documentclass{aastex}
\usepackage{emulateapj5,apjfonts,epsfig}
\def\uu {4U\,0142$+$614}
\def\src {1E\,1048.1$-$5937}

\def\axj {AX\,J1844$-$0258}
\def\rxj {1RXS\,J170849$-$400910}
\def\ee {1E\,2259$+$586}
\newcommand{\AXAF}{{\em Chandra}}

\newcommand{\EXO}{{\em EXOSAT}}
\newcommand{\e}{{\em Einstein}}
\newcommand{\xmm}{{\em XMM--Newton}}
 
\newcommand{\RXTE}{{\em R}XTE}
\newcommand {\rc}{\rm}
\newcommand{\bc}{\begin{center}}

\slugcomment{Received 2002 August 12; accepted 2002 September 27}

\shorttitle{IR VARIABILITY OF \src}
\shortauthors{ISRAEL ET AL.}

\begin{document}
\bibliographystyle{apj_noskip}

\title{The detection of variability from the candidate IR counterpart 
to the Anomalous X--ray Pulsar \src \altaffilmark{1}} 

\author{ G.L. Israel\altaffilmark{2,3}, S. Covino\altaffilmark{4}, 
L. Stella\altaffilmark{2,3}, S. Campana\altaffilmark{4,3}, G. 
Marconi\altaffilmark{5}, 
S. Mereghetti\altaffilmark{6}, R. Mignani\altaffilmark{7}, 
I. Negueruela\altaffilmark{8}, T. Oosterbroek\altaffilmark{9}, 
A.N. Parmar\altaffilmark{9}, L. Burderi\altaffilmark{2} \and 
L. Angelini\altaffilmark{10} }

\email{gianluca@mporzio.astro.it}

\altaffiltext{1}{The results reported in this Letter are partially based on 
observations carried out at ESO, La Silla, Chile (66.D-0440 and 
68.D-0350).}

\altaffiltext{2}{INAF -- Osservatorio Astronomico di Roma, V. Frascati 33, 
       I--00040 Monteporzio Catone (Roma), 
       Italy; gianluca, stella and burderi @mporzio.astro.it}

\altaffiltext{3}{Affiliated to the International Center for Relativistic 
Astrophysics}

\altaffiltext{4} {INAF -- Osservatorio Astronomico di Brera, Via Bianchi 
46, I--23807 Merate (Lc), Italy; covino and campana @merate.mi.astro.it}

\altaffiltext{5}{European Southern Observatory, Casilla 19001, Santiago, 
Chile; gmarconi@eso.org}

\altaffiltext{6}{Istituto di Fisica Cosmica G. Occhialini, CNR, 
Via Bassini 15, I--20133 Milano, Italy; sandro@mi.iasf.cnr.it}

\altaffiltext{7}{European Southern Observatory, Karl--Schwarzschildstr. 2, 
D--85748 Garching, Germany, rmignani@eso.org}

\altaffiltext{8}{Dpto. de F\'{\i}sica, Igegner\'{\i}a de Sistemas y 
Teor\'{\i}a de la Se\~{n}al, Universidad de Alicante, Apdo. de Correos 
99, E03080, Alicante, Spain; ignacio@astronomia.disc.ua.es}

\altaffiltext{9}{Astrophysics Missions Division, Research and Scientific 
Support Department of ESA, ESTEC, Postbus 299, NL-2200 AG Noordwijk, 
The Netherlands; toosterb and aparmar @rssd.esa.int}

\altaffiltext{10}{Laboratory of High Energy Astrophysics, Code 660.2, 
NASA/Goddard, Space Flight Center, MD 20771, USA; 
angelini@davide.gsfc.nasa.gov}

\begin{abstract}

We report on the detection of variability from the proposed IR
counterpart to the Anomalous X--ray Pulsar (AXP) \src\ based on
Chandra and ESO optical/IR deep observations carried out in
2001--2002.  Within the narrow \AXAF\ uncertainty region for \src\ we
found only one relatively faint ($J$=22.1$\pm$0.3, $J$--$Ks$=2.4)
source, while the recently proposed IR counterpart was not detected
down to a limiting $Ks$ magnitude of $\sim$20.7 (3$\sigma$ confidence
level). This implies a remarkable IR brightening of this object,
$\Delta\,Ks$$>$1.3, on a timescale of about 50 days. Although our
knowledge of the IR properties of AXPs is rather limited (there is
only another source, \ee, for which IR variability has been detected),
the observed IR variability of the proposed counterpart strengthens
its association with \src. Our results make the IR (and presumably
optical) variability a likely common characteristic of AXPs, and
provide new constraints on this class of objects.

\end{abstract}


\section{INTRODUCTION}
After more than 20 years since the discovery of X--ray pulsations from
the prototype source \ee, the {\rc current observational properties of
AXPs are not yet conclusive to unambiguously assess their nature}.
There are currently five confirmed members of the AXP class plus two
candidates.  Although we can be reasonably confident that AXPs are
magnetic rotating neutron stars (NSs), their energy production
mechanism is still uncertain. It is also unclear whether they are
solitary objects (either low magnetised or magnetars) or are in binary
systems with very low mass companions (for a review see Israel et
al. 2002; Mereghetti et al 2002 and references therein).  Different
production mechanisms for the observed X--ray emission have been
proposed, involving either accretion or the dissipation of magnetic
energy. The recent detection of X--ray bursts from \ee\ and {\rc \src}
has strengthened the possible connection of AXPs with Soft Gamma--ray
Repeaters (SGRs; Kaspi \& Gavriil 2002; {\rc Gavriil et
al. 2002}). Moreover the detection of relatively large pulsed fraction
pulsations in the optical flux of \uu\ (at the same period of X--ray
pulses) seems to rule out the X--ray reprocessing and favors models
based on isolated NSs (Kern \& Martin 2002). {\rc The magnetar model,
originally proposed by Thompson \& Duncan (1993) to explain the
properties of SGRs, appears currently to be fairly succesful at
interpreting these properties of AXPs.}

As done in the past in the case of High and Low Mass X--ray binary
system (HMXBs and LMXBs), finding the optical/IR counterparts to AXPs
might give the key to definitively assess the nature and the emission
mechanisms of AXPs. The lack of any known distinctive property of AXPs
in the optical/IR bands, makes the X--ray positional accuracy at the
(sub)arcsec level a fundamental preliminary step to carry out very
deep searches for potential optical/IR counterparts. Moreover a high
X--ray spatial resolution can be exploited to search for small scale
diffuse emission or the presence of structures in the vicinity of the
sources.

\src\ has been extensively studied in the X--ray band since its
discovery in 1979 (Seward et al. 1986). Previous searches for an
optical counterpart gave negative results (Seward et al. 1986;
Mereghetti et al. 1992; Wang \& Chakrabarty 2002a). Recently, a
relatively faint IR counterpart ($J$=21.7; $J$--$Ks$=2.3) to \src,
with unusual IR colors, has been proposed (Wang \& Chakrabarty 2002b; 
hereafter WC02).

In this paper we report on the early results obtained for \src\ as a
part of a joint ESO/Chandra large project aimed at the identification
and study of the optical/IR counterparts of AXPs in the southern sky,
and the study of the spatial distribution of X--ray emission from
AXPs (results concerning \AXAF\ observations of \rxj\ and \axj\ will
be reported elsewhere).

\section{\AXAF\ OBSERVATIONS}

The field including the X--ray position of \src\ was observed during
\AXAF\ Cycle 2. Previously determined positions of the source had 
limited accuracy; 15\arcsec\ for \EXO\ (Mereghetti et al. 1992) and
$\sim$4\arcsec\ for \xmm\ (Tiengo et al. 2002). The source was observed
with the High Resolution Imager (HRC--I; Zombeck et al. 1995) in order to 
obtain the most accurate X--ray position. \src\
was observed with \AXAF\ on 2001 July 4 for an effective exposure time
of 9870\,s. Data were reduced with CIAO version 2.2 and analysed with
standard software packages for X--ray data (CIAO, XIMAGE, XRONOS,
etc.). The observation was carried out with a nominal aspect solution
and latest calibration files were used. Four different detection
algorithms were used: the CIAO {\tt wavdetect} and {\tt celldetect}
tool, the XIMAGE sliding cell, and the wavelet--based {\tt pwdetect}
software developed and optimised for \AXAF\ images (Damiani et
al. 1997).  Based on a statistical study performed on several
point--like sources detected within the field of view of public HRC--I
archival data with known optical counterparts, we found a difference
up to 0\farcs2 in coordinates among the four different detection
methods. In the following, we refer to the coordinates obtained with 
the latter method (we assumed that differences in the coordinates among
the four methods are purely statistical).
\begin{center}
\includegraphics[scale=0.32]{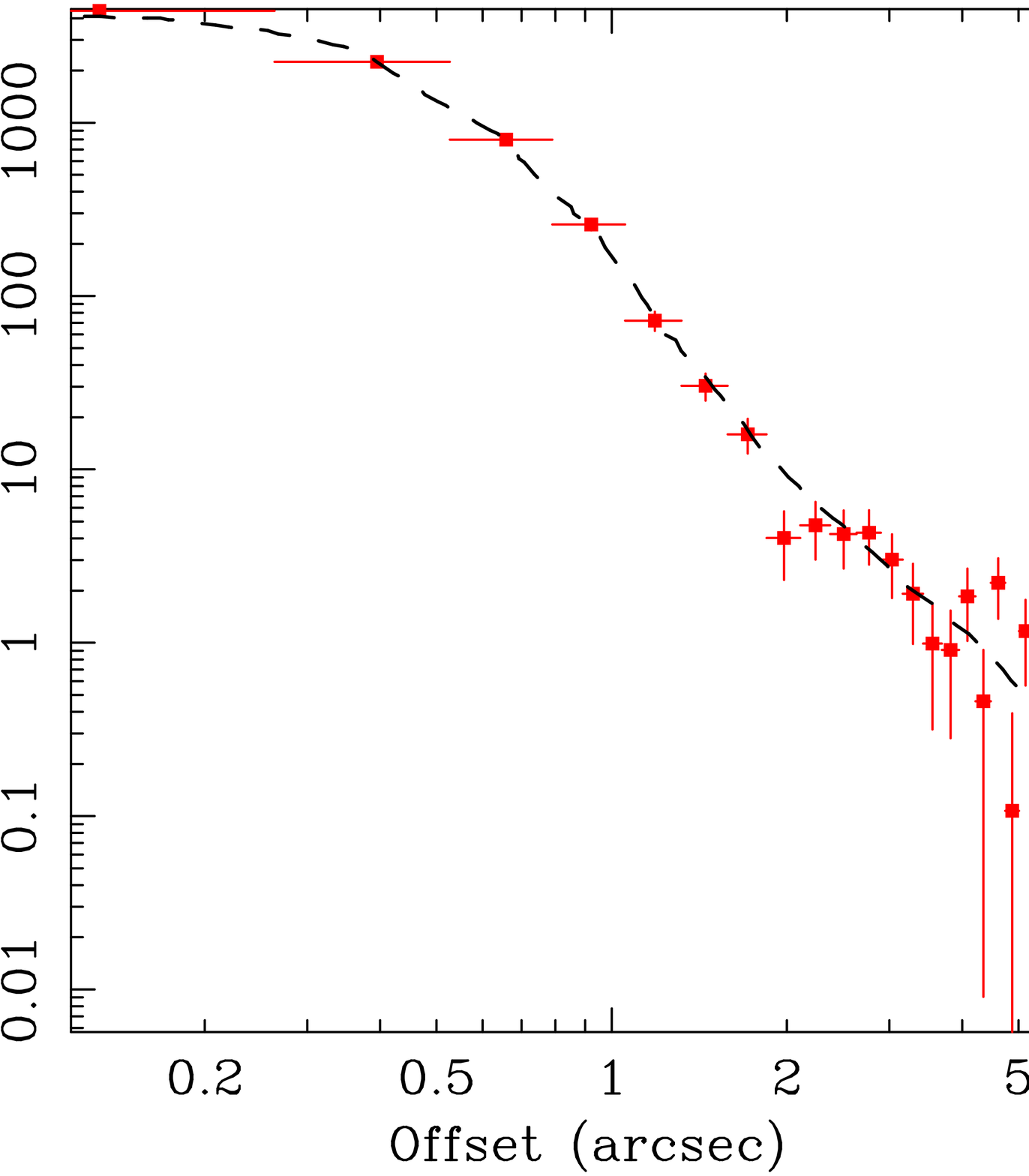}
\figcaption{Radial surface brightness profile 
of the \AXAF\ HRC--I emission from a circular region of 7\arcsec\
radius around \src. The dashed line represents the simulated
\AXAF\ point spread function for a point--like on--axis source.}
\end{center}

Only one source was detected in the HRC--I. This
has the following coordinates: R.A. = 10$^{\rm h}$50$^{\rm m}$07\fs12,
DEC. = --59\arcdeg53\arcmin21\farcs37 (equinox J2000), with an
uncertainty circle radius of 0\farcs7 (90\% confidence level). Note
that the \AXAF\ position is not consistent with that obtained with
\xmm\ (Tiengo et al. 2001; R.A. = 10$^{\rm h}$50$^{\rm m}$06\fs3,
DEC. = --59\arcdeg53\arcmin17\arcsec; equinox J2000, 90\% confidence
level circle radius of 4\arcsec). However, inspection of the attitude
data during the \xmm\ observation, showed that an error radius of
about 9\arcsec\ is likely more realistic {\rc (Tiengo 2002, private
communication)}. Photon arrival times were extracted from a
circular region centered on the previous reported position with a
radius of 1\farcs5, including more than 90\% of the source
photons. These were corrected to the barycenter of the solar system
and timing analysis applied (CIAO {\tt fxbary} tool). A coherent
pulsation at a period of 6.4529$\pm$0.0001\,s was detected confirming
that the detected source is indeed the AXP \src. The latter value
falls within the expected period interval extrapolated from the more
accurate timing analysis obtained from a long-term \RXTE\ monitoring
program of \src\ (Kaspi et al. 2001; we assumed the maximum and
minimum \.P reported in their Table\,1).

One of the aims of the \AXAF\ HRC--I observations of AXPs, was to
study the radial profile of the X--ray emission from the AXPs with the
unprecedented spatial resolution offered by the HRC--I. Figure\,1
shows the spatial profile we obtained for the (background subtracted)
X--ray emission from a region of 7\arcsec\ radius centered on the
above reported source position. The radial profile is in good
agreement with the expected \AXAF\ point spread function (solid line 
simulated with {\tt MARX}; Wise et al. 1997).

\section{ESO OBSERVATIONS} 

Over the last two years we carried out deep optical and IR
observations of the field including the original \e\ position of \src.
These data were obtained from the 3.5-m New Technology Telescope (NTT,
La Silla, Chile) equipped with the {\tt SUperb Seeing Imager-2}
(SUSI2; 2 CCDs of 2048$\times$4096 pixels; pixel scale of 0\farcs16
and 5.5\arcmin$\times$5.5\arcmin\ field of view) in the optical band,
and with {\tt Son OF Isaac} (SOFI; Hawaii HgCdTe 1024$\times$1024
array; 0\farcs292 pixel scale and 4.9\arcmin$\times$4.9\arcmin\ field
of view) in the near--IR band.

Optical Johnson filter deep images ($BVRI$) were first carried out
with {\tt SUSI2} in service mode on 2001 March 20 with the following
effective exposure times: 2700\,s ($B$), 4350\,s ($V$), 2400\,s ($R$)
and 3400\,s ($I$). Weather conditions were good with seeing of about
1\arcsec. Standard reduction packages were applied to the optical data
({\tt DAOPHOT\,II}; Stetson 1987) to obtain the photometry of each
stellar object in the images.  Limiting magnitudes of 25.5 ($B$),
25.5 ($V$), 24.5 ($R$) and 23.5 ($I$) were reached.

$JH$ images were initially carried out with {\tt SOFI} in service mode
on 2001 May 29; 2700\,s (each filter) of effective exposure time.
Unfortunately, during observations clouds were present and the seeing
was poor (1\arcsec--1\farcs5) yielding a limiting magnitude (3$\sigma$
confidence level) of 22.5 ($J$) and 21 ($H$). SOFI $JHKs$ images
were obtained with better weather (photometric) and seeing (0\farcs6)
conditions on 2002 February 19. Final exposure times were of 3600\,s
($J$), 4200\,s ($H$) and 4800\,s ($Ks$). Single 10\,s-long exposure
images were taken for each filter (5\,s-long in $Ks$) with offsets of
40\arcsec\ in order to sample and subtract the variable IR
background. Standard IR software packages were used for sky frame
subtraction and image coaddition ({\tt Eclipse} and {\tt IRDR};
Devillard 1997 and Sabbey et al. 2001).  During the latter run
limiting magnitudes of $J$$\sim$23, $H$$\sim$21.5, and $Ks$$\sim$20.7
were reached.
 
To register the \AXAF\ coordinates of \src\ on our optical/IR images,
we computed the image astrometry using as reference the positions of
stars selected from the USNO--A2.0 catalog which has an intrinsic
absolute accuracy better than 0\farcs25 (depending on magnitude and
sky position of reference stars; Assafin et al. 2001). About 40
objects with magnitudes brighter than R=14 were identified in the
optical and IR images and 
\begin{figure*}
%\begin{center}
%\includegraphics[scale=0.7]{f2e.eps}
\psfig{figure=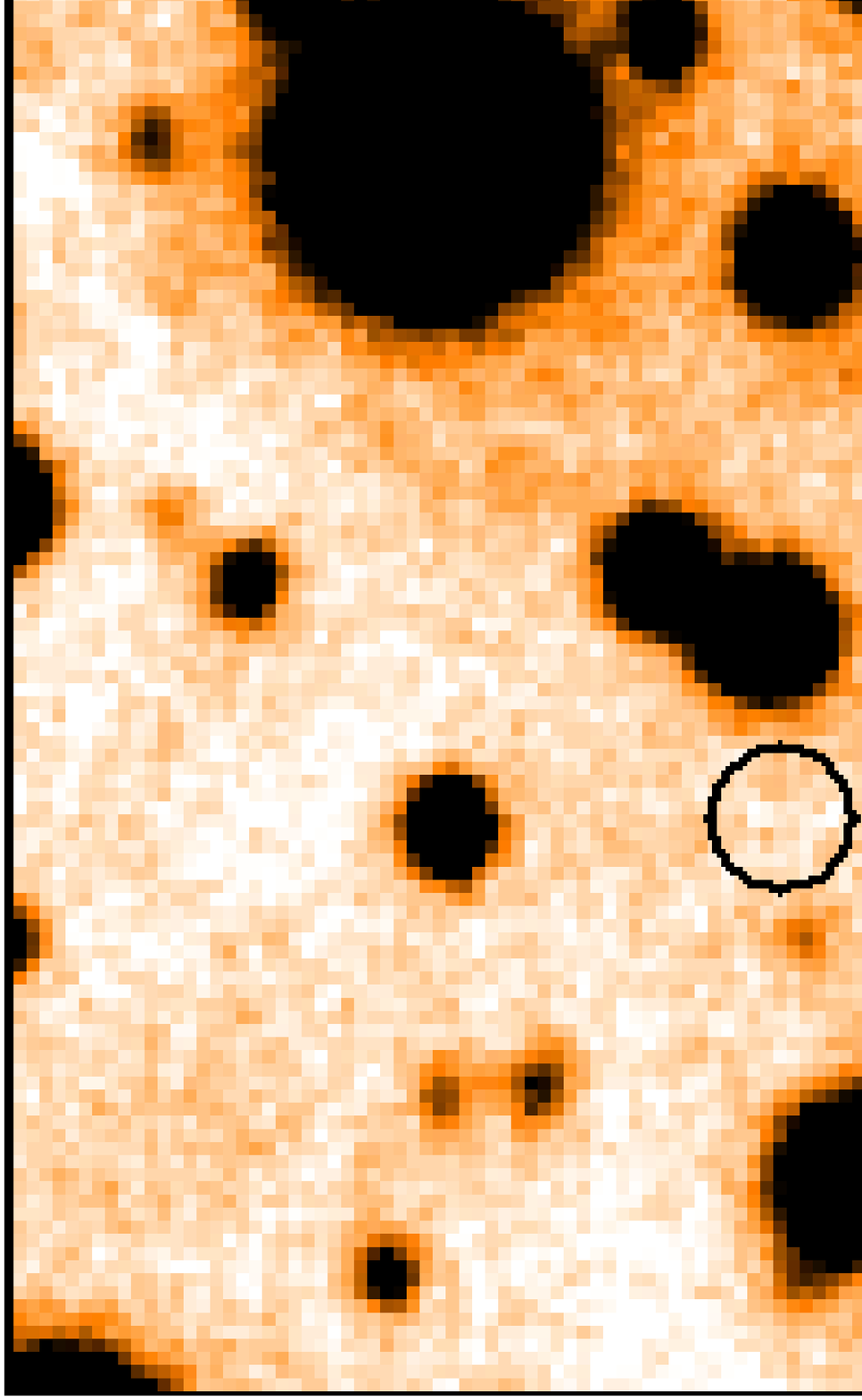,height=8.5cm}
\caption{Optical $I$ band (left panel) and
near--IR $Ks$ band (right panel) images of the region including the
position of \src, taken from the 3.5-m ESO NTT. The \AXAF\ uncertainty
circles are also shown: the thick solid line circle has radius
0\farcs8 and represents the position reported in this paper, while the
thin one is the WC02 position with a radius of 1\arcsec. The faint 
object marked with {\tt X3}, {\rc which lies on the upper
edge of the circles}, is the only one consistent with the X--ray
position. The square represents the position expected of {\tt X1},
proposed as the IR counterpart to \src\ by WC02. Coordinates are RA
(h\,m\,s) and Dec. ($^o$ \arcmin\ \arcsec; equinox J2000).}
\end{figure*}
used for astrometric calibrators.  The pixel
coordinates of the reference stars were computed by a two
dimensional Gaussian fitting procedure and transformation from pixel
to sky coordinates was then computed using the Graphical Astronomy and
Image Analysis Tool ({\tt GAIA}.\footnote{See
http://star-www.dur.ac.uk/$\sim$pdraper/gaia/gaia.html}). After taking
into account the uncertainties of the source X--ray coordinates
(0\farcs7), the rms error of our astrometry (0\farcs15) and the
propagation of the intrinsic absolute uncertainties on the USNO--A2.0
coordinates (0\farcs25), we estimated the final accuracy to be attached
to the \src\ position as about 0\farcs8. Figure\,2 shows a region of
20\arcsec$\times$20\arcsec\ around the \src\ position in the $I$ and
$Ks$ bands with the \AXAF\ uncertainty circle superimposed (90\%
confidence level).

\section{DISCUSSION}

The NTT IR data show that no object is consistent with the \AXAF\
position of \src\ in the optical band down to a $V$ limiting magnitude
of about 25.5 (see Figure\,2). On the other hand, a faint object
(marked {\tt X3} in Figure\,2) is present in the IR $Ks$ frames
($J$=22.1$\pm$0.3, $Ks$=19.7$\pm$0.3). {\rc This object is also
detected in the WC02 frames, although its magnitudes are not listed. A
visual inspection of the WC02 image suggests that the source {\tt X3}
has remained nearly unchanged in flux with respect to our frames}
(superimposing to our IR images the \AXAF\ position and
uncertainty radius reported by WC02 for \src, the object {\tt X3}
falls within the circle). {\rc Note that during the NTT IR
observations carried out on 2001, the object {\tt X3} is not
detected due to the poor seeing conditions and contamination with the
nearby bright object.} More surprisingly, we note that no object is
detected at the position of the source {\tt X1} reported by WC02 (the
position of {\tt X1} is marked with a square in Figure\,2). For
comparison, the object marked {\tt X2} in Figure\,2 has a magnitude of
$J$=20.63$\pm$0.06, $H$=19.73$\pm$0.15 and $Ks$=19.0$\pm$0.2
(consistent, within 1--1.5$\sigma$, with the values reported by
WC02). {\rc As the limiting magnitudes of our IR frames are well
above the values measured by WC02 for {\tt X1}, we conclude that the
source was considerably fainter during our observations}. This IR
brightening ($\Delta Ks>$1.3; {\rc about a factor of 3 in flux}),
displayed by the source on a timescale of about 50 days (the time
interval between our last IR images and those of WC02) further
strengthens the link between the IR source and \src. Moreover, the
chance probability of finding a highly variable IR source within a
circle of 0\farcs8 is negligible.

An IR counterpart was recently found for the prototype of the AXP
class, \ee, which also showed IR variability at a level of $\Delta
Ks\sim$1.3 (Kaspi et al. 2002). This IR variability {\rc might be
associated to the short--term X--ray bursts (similar to those
displayed by SGRs) observed from the source three days before the IR
observations} (Kaspi \& Gavriil 2002). Recently, short X--ray bursts
were detected in the X--ray flux of \src\ (Gavriil et al. 2002). All
these findings together, suggest {\rc a close similarity of \src\ and
\ee, such that the IR counterpart of the former might be expected to
be variable. For this reason {\tt X1} is the most likely IR
counterpart to \src, even though source {\tt X3} is included in the
\AXAF\ uncertainty circle and its color ($J$--$Ks$=2.4) is within the
value interval predicted for AXPs (see also Figure \,2 of WC02)}.

{\rc The fact that \src\ showed two X--ray bursts close to each other,
implies that many other bursts might have been missed due to the
sparse X--ray monitoring with the \RXTE. Moreover since our
non--detection of the IR candidate predates the WC02 detection, it
would appear more natural to associate (if at all) the IR variability
to some X--ray activity that took place between the two IR campaigns
(as opposed to the few second--duration X--ray bursts that was
detected 4 months earlier). Finally we note that the X--ray/IR
observations of \ee\ are such that we do not know whether the detected
IR brightening follows or precedes that in the X--ray band. Future
multi--wavelength monitoring programs will probably allow us to better
understand these points.}

The non--detection of {\tt X1} in our IR frames implies that the
``quiescent'' (or low state) IR emission level of \src\ is still
undetected. {\rc Correspondingly, it cannot be ruled out yet that the
spectral flattening proposed by WC02 might result from
wavelength--dependent variability}. Similarly, the measured IR fluxes
for the counterpart cannot be used to study the overall (from X--ray
to IR) energy spectrum of \src, since they reflect a different energy
distribution than the quiescent (or low state) one.  Deeper IR
observations are needed to determine the quiescent emission state
of this source.

IR variability might represents a new, fairly common, characteristic
of AXPs (and possibly SGRs), and potentially a diagniostic of their
nature. We note that, so far, very little is known on the expected
optical/IR emission from a magnetar. Both models based on magnetar and
accretion from ``fossil'' disk have difficulties in accounting for
transient {\rc and/or variable} emission {\rc at wavelengths shorter
than those in the X--ray band (Thompson \& Duncan 1996 and references
therein; Chatterjee \& Hernquist 2000 and references therein)}. {\rc
On the other hand, the analogy with the X--ray bursts of LMXBs,
suggests that the effects of the X--ray burst of AXPs might propagate
towards longer wavelengths (e.g., via reprocessing)}. It is somewhat
intriguing that at least the values of $\Delta Ks$ and colors of the
IR counterpart of \ee\ and \src\ are similar to those displayed by
some classes of accreting cataclysmic variables and low mass X--ray
binaries hosting a neutron star. 

The superb \AXAF\ HRC--I spatial resolution allowed us to look for
extended X--ray emission around the pulsar; {\rc this might result
from an X--ray nebula powered by the AXPs as well as a scattering
halo.} A first detailed study was carried out for \ee, where emission
beyond 4\arcsec\ and extending up to 100\arcsec\ was found (Patel et
al 2001). However in this case the source is embedded in the diffuse
emission from the supernova remnant G109.1--1.0, and it is very
difficult to disentangle this component from that (possibly present)
around the AXP. The present \AXAF\ HRC--I data are by far the best
spatial resolution observation carried out for an AXP and no
significant diffuse X--ray emission was found beyond $\sim$2\arcsec.

\acknowledgments{ This work is supported through CNAA, ASI, CNR and 
Ministero dell'Universit\`a e Ricerca Scientifica e Tecnologica
(MURST--COFIN) grants. The authors thank Olivier Hainaut, Leonardo
Vanzi and the NTT Team for their kind help during ESO
observations. {\rc We thank Fotis Gavriil for a careful reading of an
early version of the paper.}}

\end{document}